## ESTUDIOS / *RESEARCH STUDIES*

# BPMS for management: a systematic literature review

Alicia Martín-Navarro*; María Paula Lechuga Sancho*, José Aurelio Medina-Garrido*

*INDESS (University Research Institute for Sustainable Social Development). University of Cadiz.
E-mail: alicia.martin@uca.es | ORCID iD: https://orcid.org/ 0000-0002-9443-6491
E-mail: paula.lechuga@uca.es | ORCID iD: https://orcid.org/ 0000-0003-2340-7615
E-mail: joseaurelio.medina@uca.es | ORCID iD: https://orcid.org/ 0000-0003-3120-6426

### 

**Abstract:** The aim of this paper is to carry out a systematic analysis of the literature to show the state of the art of Business Processes Management Systems (BPMS). BPMS represents a technology that automates business processes connecting users with their tasks. For this, a systematic review of the literature of the last ten years was carried out, using scientific papers indexed in the main databases of the knowledge area. The papers generated by the search were later analysed and filtered. Among the findings of this study, the academic interest and the multidisciplinary nature of the subject, as this type of studies have been identified in different areas of knowledge. Our research is a starting point for future research eager to develop a more robust theory and broaden the interest of the subject due its economic impact on process management.

**Keywords:** information system; BPM; business process management; workflow system; systematic literature review.

Alicia Martín-Navarro, María Paula Lechuga Sancho and José Aurelio Medina-Garrido

## 1. INTRODUCTION

Traditionally, business management is based on a functional model that breaks down the Organizational structure in individual departments, this means that each department has a separate agenda from the rest of the functional areas, and different responsibilities and power (Coulson et al., 2010). The main problem with traditional management and its functional hierarchy lies in the development of barriers between individual departments (Robson and Ullah, 1996). However, modern organizations choose to focus on processes, this means focusing on customers (Reijers, 2006). In this sense, processes cross departmental boundaries and organizations become more horizontal. A process-oriented organization usually applies the concept of business process management or *Business Process Management* (hereinafter BPM) (Kohlbacher, 2010). BPM is a way of managing focused on aligning all aspects of an organization, based on an analysis, control and maintenance of its business processes. It wants the organization toGain in effectiveness and efficiency and sWhat is the goal? that the company continuously improves its processes (Malaurent & Avison, 2016).

Since the 1980s, BPM has been an intensely discussed topic in the field of information systems research (Houy et al., 2010). In previous decades, many organizations sought to embrace technology initiatives that would enable them to make changes, manage their businesses, and improve business performance (Harmon, 2010). BPMS represents a type of *software* that allows the management of an organization's business processes (van der Aalst et al., 2003), through the design and modeling of these processes (Smith and Fingar, 2003).

The analysis of the academic literature on BPMS reveals two different perspectives in its scientific treatment, Houy et al. (2010): the management perspective and the technological perspective. The first seeks to develop and justify theories in the context of the analysis, implementation, management and use of information systems. In contrast, the technological perspective focuses on the innovation, design, and use of technological devices. Therefore, academic works on BPMS can be grouped into those that give greater importance to the technological component, i.e. design, and those that attempt to analyze BPM as an administrative theory (Houy et al., 2010; Poelmans et al., 2013; Rhee et al., 2010). This paper is framed in the second perspective, that is, it will analyze BPMS from a managerial, administrative or managerial point of view.

While the concept of BPM as a management approach was first raised in mid-2019, the concept of BPM as a management approach was first raised. In the 1990s, there is still no unanimous and accepted definition in the academic literature. around the concept and its breadth (Valverde & Talla, 2012). In this sense, and with the aim of elaborating a state of the art of the question, a reExhaustive and systematic review of the literature. The The importance of literature searches, led by reviews, is discussed by an approach known as Systematic Literature Review (SLR) (Boell & Cecez-Kecmanovic, 2014).

The aim of this paper is to analyze and classify the existing literature on BPMS from a management perspective,in order to reach new conclusions and identify research needs (Rooney et al., 2014). To this end, this work is organized as follows: In the second section, the methodology used, developed after posing the research question, is described. Subsequently, the selection of works and an analysis of the data is carried out. Finally, the main conclusions, limitations and future lines of research are determined.

## 2. METHODOLOGY

The transfer of knowledge through scientific publications is considered a fundamental element for the advancement of any science (Pérez-Anaya, 2017; Restrepo Valencia et al., 2015). This breakthrough requires researchers to know the state of the art in the subjectsstudied. In this context, it is valuable to carry out a qualitative analysis, requiring a quantitative study of the academic literature on a topic (Fernández, 1998).

Among the different techniques of literature analysis, and for the specific case of the study of the literature on BPMS, the Systematic Review of the Literature (RSL) has been chosen as the most appropriate methodological approach for two reasons. First, it is systematic, explicit, and reproducible, and is therefore perfect for identifying, evaluating, and interpreting the academic literature(Hohenstein et al., 2014). Second, it is a valid method for generating knowledgethrough the synthesis of existing articlesthat may be more relevant and of greater importance than even the most current research (Cooper, 2010). The methodology used is based on the work of Biolchini et al. (2005), Da Silva et al. (2011), K. itchenham (2004), Kitchenham et al.

(2009, 2010), Kitchenham and Charters (2007), and García-Peñalvo et al. (2015).

To carry out this RSL and following the process suggested by Kitchenham and Charters (2007), a five-stage review protocol has been developed: (1) definition of the research question, (2) design of the search strategy, (3) selection of papers and data extraction, and (4) synthesis of data (Wen et al., 2012).

**2.1. Definition of the Research Question**

The aim is to answer the following *research questions* (RQ). Specifically, for its formulation we based ourselves on the research questions previously asked in the articles by Houy et al., 2010; Kitchenham et al., 2009, 2010; Sánchez and Blanco, 2016.

- **RQ1:** What are the main works that study BPMS from a management point of view?
- **RQ2:** In which journals are these works published?
- **RQ3:** How has the evolution of the work that studies BPMS been from a management point of view?
- **RQ4:** Which countries, universities and areas of knowledge are most concerned about this type of research?
- **RQ5:** Who are the most productive authors?
- **RQ6:** What research topics are being addressed within the scope of BPMS?
- **RQ7:** What are the main reasons that lead authors to research BPMS?

**2.2. Design of the search strategy**

Search strategy involves determining search terms, literary devices, and the search process (Wen et al., 2012). The search took place in January 2017.

*2.2.1. Search terms*

The following steps were followed to define the search terms to be included in the databases and to obtain the articles for this review (Wen et al., 2012):

1. The main concepts referred to in the research question were used. In addition to the terms BPMS and *Business Process Management System*, the term *Workflow* (WFMS) was also included because it represents a similar type of information systems. Some authors have even used the terms interchangeably (Reijers et al., 2016; Pistol et al., 2015; Zanoni et al., 2014).

2. Different ways of writing the terms were identified , including abbreviations and synonyms.

3. The inclusion of keywords, previously used in relevant scientific articles, was verified .

Taking into account the three previous criteria and in order to answer the research question posed, a search was carried out focusing mainly on the coincidence of the words used with the title, keywords or abstracts of the articles located in the different selected databases (Houy et al., 2010; Kitchenham et al., 2010; Tarhan et al., 2016; Wen et al., 2012).

Keywords were searched in the databases through the following construction: BPM OR BPMS OR "Business Process Management" OR "Business Process Management System" OR WFMS OR "Workflow System" OR "Workflow Management System"

*2.2.2. Literary devices*

Five digital databases were used to search for the most relevant articles on the research topic. Specifically, the sources for the literature review were the ABI databases.*Inform*, EBSCO, IEEE *Xplore*, OTHERS *Web of Science* and *Scopus*. AdeMore potential add-on items were identifiedWhen searching in *Google Scholar*. This search procedure is widely accepted and has been used in other systematic reviews of the literature in the area of information systems (David & Han, 2004; Giunipero et al., 2008; Hohenstein et al., 2014; Houy et al., 2010; Keller and Ozment, 2009; Kitchenham et al., 2010; Soni and Kodali, 2011; Tarhan et al., 2016; Turner et al., 2010; Wen et al., 2012; Winter and Knemeyer, 2013).

*2.2.3. Search process*

In order to find out the state of academic research on BPM systems from an administrative perspective, it was decided to focus this study on indexed scientific journalsfrom 1 January 2007 to 31 December 2016. Houy et al. (2010) analysed the state of the art up to 2008. Our work is of great added value as it carries out an updated study, which also represents the continuation of the work of these authors,thus filling an important gap in the chosen topic in terms of RSL.

Alicia Martín-Navarro, María Paula Lechuga Sancho and José Aurelio Medina-Garrido

In all those databases that allowed it, it was filtered by categories and by areas of research close to the topic studied. The search for articles and the application of the filters specified above generated the identification of 2,271 works. Table I summarizes all the search results across the five databases.

**Table I.** Search results through databases

| ABI Inform | EBSCO | IEEE Xplore | ISI Web of Science | Scopus |
|---|---|---|---|---|
| 639 | 476 | 240 | 303 | 613 |

### 2.3. Selection of works

In order to select the interesting works in this study, the inclusion and exclusion criteria were applied, in addition to the evaluation of quality through the quality criteria.

#### 2.3.1. Inclusion and exclusion criteria

As authors such as Kitchenham and Brereton (2013) have already done, once the keywords have been included and the articles that would form the basis of the work for this study have been obtained, the studies were continued to be refined by applying inclusion and exclusion criteria. Specifically, and following Echeverri and Cruz (2014) and Ramírez Correa and García Cruz (2005), all scientific articles with a publication date between 2007 and 2016 and published in English were retrieved. Likewise, in order to ensure the quality of the literature retrieved, we only searched for articles published in scientific journals, as David and Han (2004) did. And as a last inclusion criterion, we searched for primary studies and not reviews, as in Guinea, et al., 2016.

In the same way, based on the works of Houy et al. (2010) and Guinea et al. (2016), the main exclusion criteria were established that led to the rejection of all those works (1) of less than four pages, (2) not referring to the research topic that is business process management, (3) that dealt with process management but did not refer to the *software* tool and, finally, (4) those jobs that study software design or engineering. On the basis of the above, all those jobs that referred to BPM *software* from the point of view of organisational management and use of the system were identified. Table II summarizes all the inclusion and exclusion criteria used for the selection of articles.

**Table II.** Inclusion and exclusion criteria

| Inclusion criteria | Exclusion Criteria |
|---|---|
| ✓ Date (2007-2016) | ✓ Duplicate posts |
| ✓ Articles in scientific journals | ✓ Articles of less than four pages |
| ✓ Language: English | ✓ No Referred AI Topic (BPM) |
| ✓ Primary Jobs | ✓ They do not allude to the *software* tool |
| | ✓ Articles that study design or engineering |

The selection of articles was carried out manually, reading the *abstract* in order to discard those works that met the exclusion criteria. After carrying out this work and applying both the inclusion and exclusion criteria, the remaining articles totalled 129 papers to which a quality assessment was subsequently applied with the main objective of guaranteeing the integrity and reliability of the information presented in the final systematic review.

#### 2.3.2. Quality assessment (quality criteria)

The primary studies obtained were of many different types: case studies, surveys, theoretical, etc. Thus, instead of using multiple instruments to evaluate the quality of the different studies, following Kitchenham and Brereton (2013) and Guinea et al. (2016), the works were classified according to the type of study and a generic set of questions was used to evaluate their rigor, credibility and relevance, and thus be able to make the final screening to obtain the final sample. This quality tool was developed by Dybå and Dingsøyr (2008) in their systematic review of software engineering, and is applicable to most studies.

Specifically, the quality of each study was assessed by classifying it according to eight different criteria listed in Table III. Each of the questions had three optional answers: "Yes", "partially" and "no". Following Wen et al. (2012), these responses were scored as follows: "Yes" = 1, "partially" = 0.5, and "no" = 0. The quality assessment of the articles is calculated by adding the scores of the answers to the previously defined questions.

As Wen et al. (2012) have done in order to be as objective as possible, the researchers extracted quality data from each primary study independently. All results were compiled and disagreements were discussed until a consensus. Only those studies with acceptable quality, i.e. with a quality score greater than 4 (50% of the perfect score) were considered for subsequent data extraction and synthesis (Guinea et al., 2016). Thus, after applying the quality criteria, a total of 102 final articles were obtained. The article selection process can be seen in Figure 1.

**Table III.** Quality Criteria

| No. | Question | Author |
|---|---|---|
| QA1 | Is this an empirical study? | Dybå and Dingsøyr, 2007 |
| QA2 | Are the objectives of the research clearly definidos? | Dybå y Dingsøyr, 2007; Kitchenham and Bereton, 2013 |
| QA3 | Is there an adequate description of the context in which Was the research conducted? | Dybå and Dingsøyr, 2007; itchenham and Bereton, 2013 |
| QA4 | Was the research method or methodology appropriate? address research objectives? | Kitchenham and Bereton, 2013 to |
| QA5 | Was the data analysis rigorous enough? | Kitchenham and Bereton, 2013 |
| QA6 | Are the results of the evaluation clearly defined? | Unterkalmsteiner et al., 2012 Kitchenham and Bereton, 2013 |
| QA7 | Are the limitations of the study explicitly discussed? | Wen et al., 2012 |
| QA8 | business community? | Wen et al., 2012; Kitchenham and Bereton, 2013 |

### 3. RESULTS AND DISCUSSION

Once the final results have been obtained, the results are analysed by year, scientific journal, country, author, university, area of knowledge, type of study, research topic, study accelerators and type of software tool.

**3.1. Number of publications per year**

As can be seen in Figure 2, since 2007 with a total of four publications, contributions to the topic have increased considerably, reaching its highest peak in 2012 with sixteen articles published. Some of the reasons that motivated the increase in academic interest in BPMS were that, in the context of a major global economic crisis, these software allowed: (1) greater control over business processes, (2) great support for critical decision-making under difficultcircumstances, (3) great agility and rapid adaptation to the necessary changes imposed by the crisis, through restructuring and continuous optimization of processes and, (4) greater strategic orientation in order to achieve both long-term and short-term business objectives. Subsequently contributions drop from seventeen to ten obtained in 2015, to increase again to thirteen publications in 2016.

**Figure 1.** Stages of the process and number of final articles

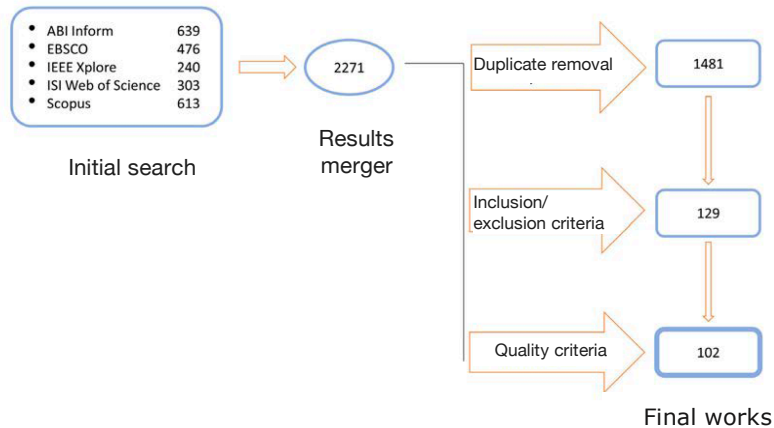

**Table IV.** Contributions per journal

| Journal Title | Nº of publications | % |
|---|---|---|
| *Business Process Management Journal* | 41 | 40.20 |
| *International Journal of Information Management* | 4 | 3.92 |
| *Information Technology and Management* | 3 | 2.94 |
| *Communications of the Association for Information Systems* | 2 | 1.96 |
| *Economic and Business Review for Central and South-Eastern Europe* | 2 | 1.96 |
| *Economic Research- Economic Research* | 2 | 1.96 |
| *Information Resources Management Journal* | 2 | 1.96 |
| *Information Systems and E-Business Management* | 2 | 1.96 |
| *International Journal of Production Economics* | 2 | 1.96 |
| *Journal of International Technology & Information Management* | 2 | 1.96 |
| *Knowledge and Process Management* | 2 | 1.96 |
| *Total Quality Management and Business Excellence* | 2 | 1.96 |
| Other | 36 | 35.29 |

**Figure 2.** Number of publications per year

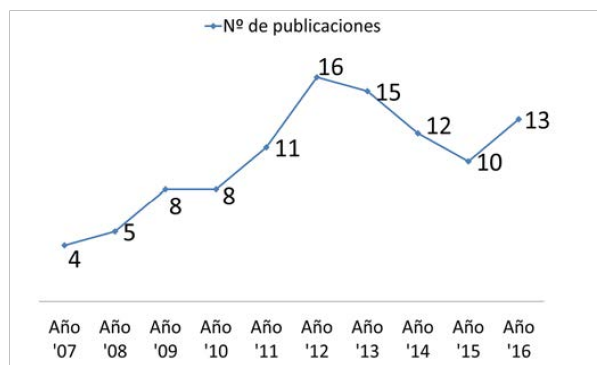

### 3.2. Contributions by scientific journal

Table IV shows the data on the number of papers published according to the journal and their percentage with respect to the total of 102 definitive papers. As can be seen, most of the papers have been published in the *Business Process Management Journal.* The

Articles published in this journal represent 40.20% of the total scientific production of this analysis. The rest of the journals publish two articles or only one, in the case of those added in the "other" section. There are 36 journals that include only one paper and represent 35.29% of the total number of documents published.

### 3.3. Contributions by country

Considering the country of origin of the first author, there are 32 countries that produce research related to the topic of BPMS with the characteristics set by this systematic review and in the last 10 years. Among those countries that have shown the greatest concern about BPM systems are Australia, Germany, the United States and Brazil. Australia is the leading country in the contributions of works related to the object of this study with 14 contributions, 13.73% of the total contributions. This was followed by Germany and the United States with 10 works, each with a representation of 9.80%. Subsequently, Brazil's production represents 6.86% and both the United Kingdom and the Netherlands contribute 6 documents, each of these countries representing 5.88% of production. The rest of the countries and their contributions are shown in Table V.

### 3.4. Contributions by author

This section lists the names of those authors who are actively involved in research on BPM tools from a management perspective, and who have published in the last decade. A total of 228 authors participated in the 102 articles obtained in this systematic review. Table VI lists the thirteen authors who have written the most articles in this type of study, with more than three articles each. The most productive researcher in the field of research is Jan *Recker* who has participated in 9 articles, followed by Jan vom Brocke with 6 publications and Bjoern Niehaves with 5. Next, in the table, three authors (J. Becker, R. Plattfaut and T. Schmiedel) appear with four publications each. And to complete the fourteen, the rest of the authors, who are seven (Y. L. Antonucci, R. Batenburg, R. J. Goeke, O. Marjanovic, S. I. of Padua, M. I. Štemberger and P. Trkman), have contributed three publications each. Of the 228 authors, seventeen others have contributed two articles each and the rest, which are 198 researchers, have published only once in the 102 articles that make up this study.

**Table V.** Contributions by country

| Country | number | % | Country | number | % |
|---|---|---|---|---|---|
| Australia | 14 | 13.73 | South Africa | 2 | 1.96 |
| Germany | 10 | 9.80 | Switzerland | 2 | 1.96 |
| USA | 10 | 9.80 | Saudi Arabia | 1 | 0.98 |
| Brazil | 7 | 6.86 | Austria | 1 | 0.98 |
| Netherlands | 6 | 5.88 | China | 1 | 0.98 |
| United Kingdom | 6 | 5.88 | Colombia | 1 | 0.98 |
| Liechtenstein | 5 | 4.90 | Croatia | 1 | 0.98 |
| Slovenia | 4 | 3.92 | Denmark | 1 | 0.98 |
| Malaysia | 4 | 3.92 | Dubai | 1 | 0.98 |
| Poland | 4 | 3.92 | Spain | 1 | 0.98 |
| Belgium | 3 | 2.94 | Holland | 1 | 0.98 |
| Italy | 3 | 2.94 | Norway | 1 | 0.98 |
| Iran | 2 | 1.96 | Russia | 1 | 0.98 |
| Lithuania | 2 | 1.96 | Slovenia | 1 | 0.98 |
| Czech Republic | 2 | 1.96 | Sweden | 1 | 0.98 |
| Romania | 2 | 1.96 | Taiwan | 1 | 0.98 |

**Table VI.** Contributions by author

| Author | Articles | number | % |
|---|---|---|---|
| Recker, J. | [64] [71] [72] [73] [74] [75] [85] [86] [97] | 9 | 3.95 |
| Vom Brocke, J. | [12] [85] [86] [97] [98] [99] | 6 | 2.63 |
| Niehaves, B. | [57] [58] [59] [60] [61] | 5 | 2.19 |
| Becker, J. | [59] [60] [61] [81] | 4 | 1.75 |
| Plattfaut, R. | [58] [59] [60] [61] | 4 | 1.75 |
| Schmiedel, T. | [85] [86] [97] [99] | 4 | 1.75 |
| Antonucci, Y. L. | [7] [31] [32] | 3 | 1.32 |
| Batenburg, R. | [21] [22] [70] | 3 | 1.32 |
| Goeke, R. J. | [7] [31] [32] | 3 | 1.32 |
| Marjanovic, O. | [47] [48] [49] | 3 | 1.32 |
| Pádua, S. I. D. de | [3] [52] [87] | 3 | 1.32 |
| Štemberger, M. I. | [13] [50] [94] | 3 | 1.32 |
| Trkman, P. | [50] [90] [97] | 3 | 1.32 |

**Table VII.** Contributions by University

| University | Articles | number | % |
|---|---|---|---|
| Queensland University of Technology | [49] [71] [72] [73] [74] [75] [78] | 7 | 10.14 |
| University of Liechtenstein | [85] [86] [88] [97] [98] [99] | 6 | 8.70 |
| University of Ljubljana | [13] [14] [41] [90] [94] | 5 | 7.25 |
| University of São Paulo | [8] [52] [62] [79] [87] | 5 | 7.25 |
| University of Münster | [12] [58] [59] [60] | 4 | 5.80 |
| University of Applied Sciences Utrecht | [21] [22] [70] | 3 | 4.35 |
| Widener University | [7] [31] [32] | 3 | 4.35 |
| Asia Pacific University of Technology and Innovation | [54] [55] | 2 | 2.90 |
| La Trobe University | [17] [20] | 2 | 2.90 |
| Tarbiat Modares University | [4] [69] | 2 | 2.90 |
| The University of Sydney | [47] [48] | 2 | 2.90 |
| University of Twente | [2] [53] | 2 | 2.90 |
| University of Warsaw | [15] [28] | 2 | 2.90 |
| Vilnius University | [83] [84] | 2 | 2.90 |

### 3.5. Contributions by university

Considering the university to which the first author of each article is attached, a total of 69 universities are represented in this study. Table VII shows the fourteen universities that produce two or more papers. The Australian *Queensland University of Technology* is the university that produces the most publications with a total of seven articles, representing 10.14% of the total scientific production selected. *The University* of Liechtenstein, in the Principality of Liechtenstein, has six publications, with a percentage of 8.70%. Two universities, the University of *Ljubljana and the* University of *São Paulo*, in Slovenia and Brazil, respectively, obtain 7.25% of scientific production with five papers each, and in Germany, the *University of Münster*, it has published four papers with a representation of 5.80%. *University of Applied Sciences Utrecht*, in the Netherlands, and *Widener University*, in the United States, with three articles account for 4.35%. And finally, Table VII shows seven more universities with a participation rate of 2.90% and two publications each. In addition to the fourteen centers shown, 55 more universities have been found that have contributed one publication each, these 55 universities represent 79.71% of the total.

### 3.6. Contributions by area of knowledge

Table VIII shows the most prolific areas of knowledge. In particular, a total of 11 different areas or departments have been identified. The area of Information Systems Management is the one that generates the highest number of publications compared to the rest of the areas, with 39 articles, which represents 38.24% of the scientific production. As can be seen in Table VIII, the next area of knowledge with the highest number of published works is the area of Business Organization. This department has produced a total of 23 articles, generating 22.55% of all scientific production. Another outstanding department is that of Computer Languages and Systems, which, with 9.80%, publishes 10 articles. Systems Engineering and Automation and Financial Economics and Accounting are the next two areas with eight and six jobs respectively, representing 7.84% for the former and 5.88% for the latter. Publishing a total of five articles, in sixth place is the Department of Computer Science and Artificial Intelligence with 4.90%. With 3.92%, the area of knowledge of Manufacturing Process Engineering is responsible for four of the articles chosen in this systematic literature review. The remaining four areas of knowledge are applied economics, with three articles, environmental technologies, with two, and applied mathematics and construction engineering with one paper each, representing 2.94% and 1.96% for the first two and 0.98% for the last two.

Table VIII. Contributions by Knowledge Area

| | Knowledge area | number | % |
|---|---|---|---|
| | Information Systems Management | 39 | 38.24 |
| 650 | Business Organization | 23 | 22.55 |
| 570 | Computer languages and systems | 10 | 9.80 |
| 520 | Systems Engineering & Automation | 8 | 7.84 |
| 230 | Financial Economics and Accounting | 6 | 5.88 |
| 75 | Computer Science and Artificial Intelligence | 5 | 4.90 |
| 515 | Engineering of manufacturing processes | 4 | 3.92 |
| 225 | Applied Economics | 3 | 2.94 |
| 790 | Environmental Technologies | 2 | 1.96 |
| 595 | Applied mathematics | 1 | 0.98 |
| 510 | Construction Engineering | 1 | 0.98 |

Table IX. Type of study

| SCOPE OF RESEARCH | Articles | % | TYPE OF STUDY | Articles | % |
|---|---|---|---|---|---|
| Empirical research | 72 | 70,59 | Survey | 34 | 33.33 |
| | | | Case Study | 36 | 35.29 |
| | | | Experiment | 2 | 1.96 |
| Desk research | 30 | 29,41 | Literature Review | 15 | 14.71 |
| | | | Conceptual | 15 | 14.71 |

### 3.7. Type of study

Academic research divides work into two main areas, empirical research and documentary research (Bhosale & Kant, 2016). Within empirical research, which is evidence-based, Papers have been classified into case studies, questionnaires or experiments. And inside the invessArticles are divided into literature reviews or concept papers. Table IX shows the classification by type of study.

### 3.8. Contributions by research topic

Business process management is a discipline that encompasses the analysis, modelling, implementation, execution, control and optimisation of processes (Minonne & Turner, 2012). All of these aspects represent the stages of the life cycle of the BPMS. In this section, the 102 Articles analyzed in this systematic review of the literature on 7 different research topics (the first 4 referring to the life cycle): (1) modeling, (2) implementation, (3) use of the system, (4) evaluation, (5) integration, (6) training and (7) optimization.

The results of the analysis of the articles (summarized in Table X) show that in 32 of the 102 articles analyzed, the authors investigate the evaluation and success of the system, which represents 31.37% of the total number of articles. The next most mentioned topic is implementation, specifically studied in 20 studies with a representation of 19.61%. The use of the system is the third most studied topic, with 18 articles and one 17.65% incidence. Eighteen articles investigate culture is present in 6 articles (5.88%). The determination of the success factors in the implementation of the systems has been studied in 6 articles (5.88%). Collaboration between companies is present in 5 of the articles studied (4.90%). Measuring the system's performance, its level of maturity, its process orientation, and the competitive advantage that can be generated by the process modelling, accounting for 14.71%. Integration and training are topics that obtain an incidence of 8.82% and 6.86% of the total number of articles, specifically there are 9 and 7 publications that respond to both topics. And finally, a single job refers to the optimization of processes, which means an incidence of 0.98%.

**Table X.** Classification by Research Topic

| Topic | Articles | number | % |
| --- | --- | --- | --- |
| Evaluation | [3] [5] [7] [9] [10] [11] [17] [23] [25] [28] [30] [31] [37] [41] [50] [53] [57] [59] [61] [64] [79] [80] [82] [83] [84] [85] [86] [90] [94] [96] [100] [101] | 32 | 31.37 |
| Implantation | [1] [4] [13] [14] [16] [21] [22] [27] [32] [35] [39] [62] [63] [70] [76] [89] [97] [99] [102] | 20 | 19.61 |
| Use | [2] [8] [12] [15] [20] [33] [36] [40] [44] [51] [52] [56] [58] [59] [78] [87] [88] [98] | 18 | 17.65 |
| Modelling | [18] [19] [24] [29] [34] [43] [46] [65] [67] [69] [71] [72] [73] [81] [95] | 15 | 14.71 |
| Integration | [26] [42] [47] [48] [66] [68] [77] [91] [92] | 9 | 8.82 |
| Formation | [38] [49] [54] [55] [74] [75] [93] | 7 | 6.86 |
| Optimisation | [6] | 1 | 0.98 |

### 3.9. Study accelerators

In this section, articles are classified according to the different reasons that motivate authors to research BPMS. The motivators identified have been classified into 14 different categories. The "other" section includes accelerators that have been mentioned only once. These categories are summarized in Table XI.

Process improvement is the most cited reason in the different studies with a total of 17 articles (16.67% of the total). The second most interesting reason for the authors is to know the competencies and skills that both professionals and users of BPM systems have to possess for the tool to be successful. A total of 12 articles were identified with this motivator (11.76%). The capacities of the BPMS to be able to fulfill their function motivate 11 articles (10.78%). Alignment with strategy is included in 8 articles (7.84%). Organizational

**Table XI.** Study Accelerators

| Motivators | Articles | number | % |
|---|---|---|---|
| Process Improvement | [1] [6] [18] [19] [20] [25] [29] [30] [33] [34] [46] [55] [57] [69] [91] [95] [97] | 17 | 16.67 |
| Competencies & Skills | [7] [12] [21][22] [26] [31] [43] [44] [49] [72] [78] [93] | 12 | 11.76 |
| BPM Capabilities | [5] [9] [13] [16] [53] [59] [73] [75] [80] [81] [87] | 11 | 10.78 |
| Alignment with strategy | [10] [40] [51] [52] [61] [65] [68] [92] | 8 | 7.84 |
| Organisational culture | [4] [15] [83] [85] [86] [98] | 6 | 5.88 |
| Implementation Success Factors | [35] [70] [73] [76] [82] [102] | 6 | 5.88 |
| Business-to-business collaboration | [2] [27] [45] [59] [60] | 5 | 4.9 |
| Performance Measurement | [17] [37] [42] [100] | 4 | 3.92 |
| System Maturity Level | [28] [63] [67] [84] | 4 | 3.92 |
| Process Orientation | [8] [48] [53] [90] | 4 | 3.92 |
| Competitive Advantage | [47] [50] [55] [66] | 4 | 3.92 |
| *Software* Selection | [41] [79] [94] | 3 | 2.94 |
| User satisfaction | [24] [64] [101] | 3 | 2.94 |
| Process governance | [23] [32] | 2 | 1.96 |
| Other | [3] [11] [14] [36] [38] [40] [62] [74] [77] [88] [89] [96] [99] | 13 | 12.75 |

BPMS presented 4 articleseach (3.92%) in the literature analyzed. On the other hand, software selection, user satisfaction with the system and process governance have received less attention (less than 3%, considering them separately). Finally, the "Other" section includes 13 articles and each of them has a different motivator, such as, among others, the adjustment of process management with the environment, change management, implementation challenges, the quality of processes or the impact ofcreativity on business processes. These 13 articles represent 12.75% of the works that are the subject of this study.

### 3.10. Type of software tool

As previously specified, *workflow* systems have been treated in the academic literature on numerous occasions as synonyms for BPMS. For this reason, it is considered interesting to determine which of the 102 definitive works refer to one or the other. The result of the analysis determines that of the 102 articles, 95 refer to BPM systems, which represents 93.14% of the total number of articles, and only 7, with 6.86%, refer to Workflow tools ([2], [19], [39], [44], [45], [76], [91]*).* Thisresult is graphically shown in Figure 3.

**Figure 3.** Workflow vs. BPMS Systems

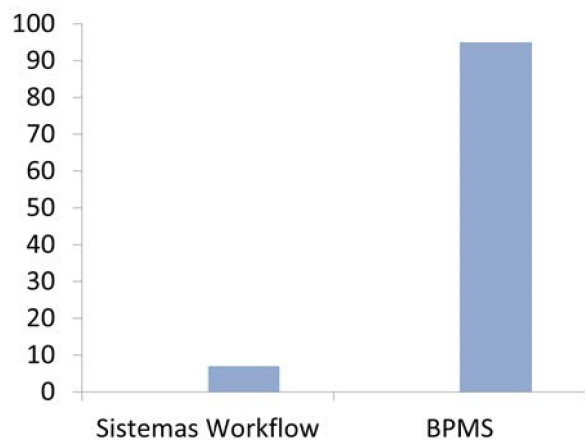

*Relevance of Articles*

To determine the relevance of the articles found, the number of citations they receive in WOS is used. The total number of citations a paper receives is the best indicator of its influence (Thomé et al., 2016). Only 13 of the 102 papers have never been cited, representing 12.74% of the articles. 52.9%, or 54 documents, received between 1 and 10 citations. Fourteen papers have been cited between 11 and 20 times, representing 13.72%. And with more than 21 citations, 17 works have been obtained with 16.67%. In short, there are 1158 citations obtained by the 102 articles retrieved in thissystematic review, of which the first ten obtained 522, which represents 45.07% of the total citations.

Table XII shows the most cited articles in this review. Most of these articles have been published in the *Business Process Management Journal*. The most cited article belongs to *J. Recker*, who in turn is the author with the most publications.

**4. CONCLUSIONS**

To meet the objective of this research, a systematic review of the literature has been carried out, following the different stages that are part of this method (Wen et al., 2012). Thus, a research question was determined, from which a search of the main scientific articles on BPMS was conducted.To do this, the search focused on a total of seven keywords that were used in five different databases. With the results obtained from this search, the exclusion and inclusion criteria were applied, and finally, in order to obtain the final articles, an evaluation was carried out through eight quality criteria. Once the article selection stage was completed, the results were analyzed and synthesized.

Academic research on BPMS from a management perspective is trending upwards, so it seems that more and more research is being done on this topic. The BPMS paradigm is published in many different types of journals, indicating that it is a multidisciplinary topic. Even so, the magazine that publishes the most articles on both topics is *Business Process Management Journal*, which is, as its name suggests, a magazine specialized in BPM. Thus, BPM research is of great interest, since, as Houy et al. (2010) also pointed out, the existence of a specialized journal, as well as specific congresses and even university titlesin this field, make BPM not a passing fad but a consolidated field of research.

Australia is the country that produces the mostscientific research in the field of BPMS, and two universities in this country, the University of Queensland and the University of Sydney, publish the most on it. Specifically, Dr. Jan Recker, who is the most prominent author in research on this type of system, is attached to the University of Queensland. On the other hand, Central European countries, such as Germany, the United Kingdom, the Netherlands or the Principality of Liechtenstein are also very involved in research on these systems, highlighting in these countries the University of Liechtenstein and the public University of Münster in Germany, and specifically Dr. Jan vom Brocke attached to the former and Dr. Björn Niehaves attached to the latter.

Another aspect to highlight is that within universities, the area of knowledge that is most committed to this type of research is the area of Information Systems Management. Another area tremendously interested in BPMS is the area of Business Organization, which studies this paradigm from a management point of view, which coincides with the focus of this work. In any case, this study identifies very different areas of knowledge such as, for example, Applied Economics, Applied Mathematics or different areas of engineering. This confirms themultidisciplinary nature of the field of BPMS.

In the review of BPMS from a management approach, within the empirical research, the case studies are practically the same as the quantitative studies through questionnaires. Among the research topics, in the same order, the evaluation of system success, implementation, system usage and modelling. Thus, researchers are very interested in determining the causes of the success of the systems and even in the development of action guides. A closer look at the articles has revealed the reasons, also called accelerators, that have prompted researchers to write about this paradigm. The one that most appeals to researchers is business process improvement. The results of the studies highlight that by improving processes, BPMS are very useful for the business world. Principally, because of the efficiency and effectiveness gains that these systems can bring to organizations. Another interesting motivator to highlight is to analyze the competencies and skills of all those people related to BPMS. This indicates the academic interest in human resources dealing with this type of system and the intention to study the competencies and skills they must possess in order to be able to apply best practices.

**Table XII.** Most Cited Articles

| Author | Year | Title | Magazine | Other |
|---|---|---|---|---|
| Recker, J. | 2010 | Opportunities and constraints: The current struggle with BPMN | *Business Process Management Journal* | 84 |
| Röglinger, M., Pöppelbuß, J., & Becker, J. | 2012 | Maturity models in business process management | *Business Process Management Journal* | 74 |
| Bai, C., > Sarkis, J. | 2013 | A grey-based DEMATEL model for evaluating business process management critical success factors | *International Journal of Production Economics* | 58 |
| McCormack, K., Willems, J., Van den Bergh, J., Deschoolmeester, D., Willaert, P., Indihar Štemberger, M. & Bosilj Vuksic, V. | 2009 | A global investigation of key turning points in business process maturity | *Business Process Management Journal* | 58 |
| Houy, C., Fettke, P., & Loos, P. | 2010 | Empirical research in business process management - analysis of an emerging field of research | *Business Process Management Journal* | 54 |
| Mutschler, B., Reichert, M., & Bumiller, J. | 2008 | Unleashing the effectiveness of process-oriented information systems: Problem analysis, critical success factors, and implications | *IEEE Transactions on Systems, Man, and Cybernetics, Part C (Applications and Reviews)* | 51 |
| Neubauer, T. | 2009 | An empirical study about the status of business process management | *Business Process Management Journal* | 42 |
| Vergidis, K., Turner, C. J., & Tiwari, A. | 2008 | Business process perspectives: Theoretical developments vs. real-world practice | *International Journal of Production Economics* | 40 |
| Ravesteyn, P., & Batenburg, R. | 2010 | Surveying the critical success factors of BPM-systems implementation | *Business Process Management Journal* | 32 |
| Vom Brocke, J., & Sinnl, T. | 2011 | Culture in business process management: A literature review | *Business Process Management Journal* | 29 |

As already explained, there are authors who deal with other tools, specifically workflow systems, such as BPMS, although the concept of BPMS is used much more frequently than the concept *of workflow*. And finally, to determine that practically all the articles extracted from this review receive some citation by other authors in WOS, although almost half of the total citations are concentrated in the ten most referenced articles.

### 4.1. Limitations and future lines of research

The results of the research should be interpreted taking into account some limitations, mainly with respect to the underlying research method. In particular, the application of content analysis does not allow the study to be exempted from a certain subjectivity. In order to avoid or minimise it, this analysis has been carried out by two different researchers, with a third researcher intervening to coordinate the solution to discrepancies or different interpretations when quantifying the information. Similarly, another limitation is the number of databases consulted and that only academic articles published in peer-reviewed journals have been searched. Perhaps, if the search is extended to a greater number of repositories and other types of publications (books, book chapters, contributions to conferences, doctoral theses or even informative works), more works could be found that confirm the findings of this research or, on the contrary, that represent a new perspective for literature in the field. In any case, we chose to include only academic articles published in journals to ensure their quality.

The contribution of this work to other researchers is twofold. On the one hand, those researchers who work in the BPM management line have an up-to-date state of the art and can know which are the current and emerging schools of knowledge, which topics are the most discussed and which have gaps that could be exploited in future lines of research. On the other hand, this work will also be very useful for those researchers and organizations related to documentation services orlibrary services of social sciences, to know what topics and areas of research within this area should be promoted or what type of journals should be acquired, which will be useful in future lines of research.

Considering future lines of research on BPMS, it is observed that the concentration of scientific production on BPMS in certain countries, authors and universities highlights the embryonic state of research on this topic in the area of management and, more specifically, in the areaof business organization.Sa. Therefore, on the one hand, it is necessary to promote this field of research internationally, in order to generate global empirical studies on BPMS that allow an international comparison and, in this way, promote the proven advance of theoretical and practical knowledge on thistype of systems. On the other hand, research should be promoted from the perspective of the area of business organization, since it could provide a broader vision of the routines to be automated and how to optimize organizational processes. This perspective would be more useful for professionals who implement BPMS with the intention of continuously improving processes. Finally, there is a significant scarcity of studies in the literature analysed that linkBPMS with the satisfaction of users who use these systems, and this is an emerging line of study in this area.

## 5. ACKNOWLEDGMENTS

This publication and research has received partial financial support from INDESS (University Research Institute for Sustainable Social Development), University of Cadiz, Spain.

**ANEXO I**